# Observation of quasi-two-dimensional Dirac fermions in ZrTe$_5$

Xiang Yuan[1,2][†], Cheng Zhang[1,2][†], Yanwen Liu[1,2], Awadhesh Narayan[3,4], Chaoyu Song[1,2], Shoudong Shen[1,2], Xing Sui[1,2], Jie Xu[1,2], Haochi Yu[1,2], Zhenghua An[1,2], Jun Zhao[1,2], Stefano Sanvito[3], Hugen Yan[1,2]*, Faxian Xiu[1,2]*

[1] State Key Laboratory of Surface Physics and Department of Physics, Fudan University, Shanghai 200433, China

[2] Collaborative Innovation Center of Advanced Microstructures, Nanjing University, Nanjing 210093, China

[3] School of Physics, AMBER and CRANN Institute, Trinity College, Dublin 2, Ireland

[4] Department of Physics, University of Illinois at Urbana-Champaign, Illinois, USA

[†] These authors contributed equally to this work.

[*] Correspondence and requests for materials should be addressed to F. X. (E-mail: faxian@fudan.edu.cn), and H. Y. (E-mail: hgyan@fudan.edu.cn).

**Abstract**

Since the discovery of graphene[1,2], layered materials have attracted extensive interests owing to their unique electronic[3] and optical[4] characteristics. Among them, Dirac semimetal, one of the most appealing categories, has been a long-sought objective in layered systems beyond graphene.[5,6] Recently, layered pentatelluride $ZrTe_5$ was found to host signatures of Dirac semimetal.[7-9] However, the low Fermi level in $ZrTe_5$ strongly hinders a comprehensive understanding of the whole picture of electronic states through photoemission measurements, especially in the conduction band.[7,10] Here, we report the observation of Dirac fermions in $ZrTe_5$ through magneto-optics and magneto-transport. By applying magnetic field, we observe a $\sqrt{B}$ - dependence of inter-Landau-level resonance and Shubnikov-de Haas (SdH) oscillations with non-trivial Berry phase, both of which are hallmarks of Dirac fermions. The angular-dependent SdH oscillations show a clear quasi-two-dimensional feature with highly anisotropic Fermi surface and band topology, in stark contrast to the 3D Dirac semimetal such as $Cd_3As_2$. This is further confirmed by the angle-dependent Berry phase measurements and the observation of bulk quantum Hall plateaus. The unique band dispersion is theoretically understood: the system is at the critical point between a 3D Dirac semimetal and a topological insulator phase. With the confined interlayer dispersion and reducible dimensionality, our work establishes $ZrTe_5$ as an ideal platform for exploring exotic physical phenomena of Dirac fermions.

Layered materials, formed by stacking strongly bonded layers with weak interlayer coupling, have drawn immense attention in fundamental studies and device applications owing to their tunability in band structures and Fermi energy.[3, 4, 11-13] Unlike other layered materials such as $MoS_2$ and BN, graphene stands out as an appealing candidate as it is featured with a linear energy dispersion and low-energy relativistic quasi-particles.[9, 14, 15] Many exotic phenomena, such as half-integer quantum Hall effect[1, 2] and Klein tunneling[16], have been realized in graphene. Along this line, extensive efforts were also devoted to explore new Dirac semimetal states in other layered systems beyond graphene.[5, 6]

Pentatelluride $ZrTe_5$ with layered orthorhombic structure has been widely studied since 1980s for the resistivity anomaly[17-19] and large thermo-power[20, 21]. For a long time, $ZrTe_5$ was considered to be a semimetal or degenerated semiconductor with a parabolic energy dispersion.[10, 22] However, a recent study[7] revealed a linear dispersion in $ZrTe_5$ bulk states along with a chiral magnetic effect, hosting the signatures of Dirac semimetal. Nevertheless, owing to the relatively low Femi level in $ZrTe_5$, a complete understanding of the band structure, especially the conduction band, remains elusive from angle-resolved photoemission spectroscopy (ARPES) measurements, which makes it challenging to confirm the existence of Dirac fermions. Meanwhile, the layered structure of $ZrTe_5$ gives rise to a weak interlayer coupling, which should release the confinement on the interlayer dispersion of Dirac fermions as in the case of cuprates.[23] The interplay between Dirac fermions and the interlayer confinement may result in intriguing physical properties yet to be explored.

Here we report the observation of massless Dirac fermions in layered ZrTe$_5$ based on two independent experiments: magneto-optics and magneto-transport. External magnetic field leads to the Landau quantization of Bloch electrons, which enables us to probe the band structure and carrier dynamics in ZrTe$_5$. A $\sqrt{B}$-dependence of inter-Landau-level resonance is observed, indicating a linear band dispersion. Owing to the high electron mobility and low Fermi level, we are able to detect the Shubnikov-de Haas (SdH) oscillations close to the quantum limit, from which a non-trivial Berry phase is obtained. Both of them are well-established signatures for ultra-relativistic quasi-particles in crystals. Furthermore, the angular-dependent magneto-transport reveals a quasi-two-dimensional Fermi surface. A striking anisotropy is witnessed in the band dispersion characteristics along different crystal orientations, and along the b-axis (the layer-stacking direction), a carrier mass heavier than that of the free electron suggests a non-linear dispersion. This conclusion is further supported by the critical evidence of the trivial Berry phase and the observed bulk quantum Hall effect (QHE). ZrTe$_5$ is theoretically analyzed to be at a critical phase between a 3D Dirac semimetal and a weak topological insulator. Reducible dimension in ZrTe$_5$ (2D & 1D) attained upon exfoliation also promises possible device applications.

Under external magnetic field *B*, the charged particles can occupy discrete orbits and form Landau levels. In classical system described by the Schrodinger equation, by ignoring the dispersion along the parallel direction, the energy of each Landau level can be strictly derived as $E_n = (n+1/2)\hbar\omega_c, n = 0, 1, 2...$ where $\omega_c = eB/m^*$, $n$, $\hbar$ and $e$ denote the cyclotron angular frequency, Landau level index, reduced Plank's constant and elementary charge, respectively. Since $E_n$ increases linearly with $B$, magneto-optics measurements are usually applied to obtain the effective mass $m^*$ of these classical quasi-particles. However, in Dirac materials, where the electrons are described by the Dirac equation, when ignoring the parallel dispersion, the energy of the relativistic quasi-particles can be derived as

$$E_n = \text{sgn}(n)\sqrt{2v_F^2 eB\hbar |n|}, n = \pm 0, 1, 2... \quad (1)$$

Here $E_n$ follows a square root relation with the magnetic field as depicted in Fig. 1a. The Fermi velocity determines the evolution of Landau level energy in Dirac systems with magnetic field. The assumption of $k_z = 0$ is valid for magnetic fields applied along the b-axis direction. Although the Landau levels have a finite dispersion in the $k_z$ direction, magneto-optical reflection occurs at $k_z=0$ where the joint density of states of the Landau levels corresponding to the transition are optimal.[24]

Landau level transitions give resonance peaks in the reflection spectra. To confirm the validity of Equation (1) in ZrTe$_5$, we measured the reflection spectra of bulk ZrTe$_5$ crystals under various magnetic fields at liquid helium temperature. Single crystal ZrTe$_5$ used in this study was grown by iodine vapor transport method which is different from flux method[25-27]. The as-grown crystals crystallized in the orthorhombic layered structured with space group $D_{2h}^{17}$.[17] The prismatic chains of ZrTe$_3$ run along a-axis and are linked by the zigzag chains of Te along c-axis. These two dimensional layers then stack along b-axis forming the bulk crystal. In order to avoid oxidation and contamination, the samples were freshly cleaved prior to experiments. During the magneto-infrared measurements, the magnetic field is parallel to the b-axis (stacking axis) of the samples. The measured reflection spectra are normalized by the spectrum at zero magnetic field. The detailed experimental setup for the magneto-infrared measurements is provided in the supplementary Fig. S1. Several reflection peaks can be well resolved in the normalized reflection spectra measured under different magnetic fields (Fig. 1b inset). The blocked grey area is originated from 60 Hz harmonics. Since these reflection maxima systematically shift towards higher energy with increasing magnetic field, we conclude the resonance coming from inter-Landau-level transitions, as determined by electric-dipole-selection rules.[28] The incident photons excite the electrons from the occupied valance band to the unoccupied conduction band (Fig. 1a) following $\Delta n = n \pm 1$. Thus, the resonance energy of inter-Landau-level transitions is given by

$$E_{\text{resonance}} = E_n + E_{n+1} = 2v_F\sqrt{eB\hbar}(\sqrt{n}+\sqrt{n+1}), n=0,1,2... \quad (2)$$

The energy ratio between the second (right) and the first (left) resonance peak (Fig. 1b inset) is 1.3, almost the same value as $(\sqrt{2}+\sqrt{3}):(\sqrt{1}+\sqrt{2})$. So the first peak is assigned to be the transition from $L_{-1}$ to $L_2$ or from $L_{-2}$ to $L_1$ (orange arrows in Fig. 1a), where $L_n$ represents the $n^{\text{th}}$ Landau level. Peaks at higher energy should correspond to higher Landau level index. Keeping this in mind, we plot the peak positions with $\sqrt{n} + \sqrt{n+1}$. As shown in Fig. 1b, for different magnetic fields, the data points are well aligned on a straight line with the y-axis offset close to zero, proving the validity of Equation (1) and (2) and the Landau index assignment. These phenomena show a striking difference from classical systems such as narrow bandgap semiconductors (as well as trivial bulk state of topological insulators), in which Landau levels are equally spaced and the resonance energy is proportional to *B*. The assignment of the Landau index quantitatively leads to a Fermi velocity of $v_\mathrm{F} = 2\times10^5$ m/s, which corresponds to the average slope of the band dispersion in $k_\mathrm{a}$-$k_\mathrm{c}$ plane.

Another key feature of massless Dirac fermions in magneto-infrared spectra,[15, 28, 29] as expressed by Equation (1), is the $\sqrt{B}$-dependence of the transition energy for a fixed Landau index. In order to analyze the evolution of the peak position with the magnetic field (Fig. 2 inset), we plotted the peak position of the normalized reflection spectrum against $\sqrt{B}$ (Fig. 2). It is evident that the peak position of one Landau level transition under different fields forms a straight line, indicating a $\sqrt{B}$-dependence of resonance energy, which is different from the *B*-dependence of a classical system. The straight lines in Fig. 2 are the linear fit to the peak positions, which point to the origin of the coordinates, consistent with Equation. We noticed a recent report with a similar field dependence for the resonance energy but with different Landau level index and Fermi level.[27] The slopes of the lines derive the same value of Fermi velocity as determined in Fig. 1b. The Landau index can be also obtained from comparing the slopes of different sets of transitions. The most prominent peaks in Fig. 1b inset are determined to be $L_{-2}$ to $L_1$ ($L_{-1}$ to $L_2$). We note that the transition for lower index ($L_0$ to $L_1$) is not observable due to the strong phonon absorption of ZrTe$_5$ when the excitation energy is lower than 200 cm$^{-1}$ (Fig. S2). No clear sign of quasi-particle gap was observed in the experimental limit (refer to Section XII in the supplementary information). Both $\sqrt{n}$ and $\sqrt{B}$ dependence in the magneto-infrared study provides a consolidate evidence for the existence of Dirac fermions in ZrTe$_5$.

Recently, 3D Dirac semimetals have been extensively studied.[30, 31] The definition of the 3D Dirac semimetal requires a linear dispersions along all the directions including the high symmetry axes in crystals.[31] To confirm the dimensionality of the observed Dirac state, it is definitely insufficient to solely examine the energy dispersion in the a-c plane by the cyclotron resonance through SdH oscillations/magneto-optic or by ARPES spectroscopy. In contrast, a direct evidence for the linear dispersion along the b-axis, such as linear $E$-$k_b$ from ARPES, $\sqrt{B}$ cyclotron energy from magneto-optical spectroscopy (a-b & b-c plane) or the extraction of a Berry phase of $\pi$ from low-temperature transport measurements (a-b & b-c plane), is required to claim the 3D feature of the massless Dirac fermions. The layered structure of ZrTe$_5$ provides an ideal platform for the study of such Dirac states.

To acquire in-depth understanding of the Dirac quasi-particles, we carried out magneto-transport measurements with rotatable field direction. As schematically illustrated in Fig. 3a, conventional six-terminal devices were prepared with a constant current applied along the a-axis of the single crystal. Similar to the previous studies,[17, 18, 32] we observed several unique features of ZrTe$_5$, such as "resistivity anomaly" and Hall sign reversal at 150K (refer to Fig. S4). The anomaly temperature is related to the Fermi level position relative to the Dirac point.[33] As determined from the Hall effect (Fig. S5), ZrTe$_5$ has an *n*-type conductivity at low temperatures. Fig. 3b exhibits the magneto-resistivity (MR) of ZrTe$_5$ with the magnetic field tilting from the b-axis to the a-axis. It is evident that the rotation of the magnetic field from perpendicular to parallel to the a-c plane causes a sharp decrease of MR. Apart from the giant anisotropic MR, we observed clear SdH oscillations as a result of Landau quantization. The oscillations can be well resolved from the MR (Fig. 3b) even with the magnetic field lower than 1T, which indicates a high carrier mobility in ZrTe$_5$, consistent with the Hall effect measurements (Fig. S5&6). The amplitude of the oscillations decreases with increasing the temperature but the oscillatory frequency remains the same (Fig. 3c).

Generally, under a perpendicular magnetic field *B*, closed cyclotron orbits follow the Lifshitz-Onsager quantization rule,

$$S_\text{F}\frac{\hbar}{eB} = 2\pi(n+\gamma) = 2\pi(n+\frac{1}{2}-\frac{\phi_\text{B}}{2\pi}) \qquad (2)$$

where $S_\text{F}$ is the cross-sectional area of Fermi surface related to the Landau index *n*, and $\phi_\text{B}$ is a geometrical phase known as Berry phase. In a Dirac system, the *k*-space cyclotron orbits enclose a Dirac point so that there exists a "zero mode" that does not shift with *B*, resulting in a non-trivial Berry phase $\phi_\text{B}=\pi$.[14, 15, 34] Furthermore $\gamma$, defined as $\gamma \equiv 1/2 - \phi_B/2\pi$, can be quantitatively determined by analyzing the intercept of Landau fan diagram from the SdH oscillations. The non-trivial Berry phase generally serves as a key evidence for Dirac fermions and has been observed in several well-known Dirac materials such as graphene[2, 14], topological insulator[35, 36], and $Cd_3As_2$[37, 38]. Here we present the Landau fan diagram in Fig. 3d, where $n$ and $1/B_n$ are extracted from the peak positions in the MR curves. We only use the MR data below 2 T in this plot since the SdH oscillations are overwhelmed by a huge shoulder-like MR background at higher fields. By performing a linear fit, the intercept on the y-axis is determined to be $0\pm0.04$, corresponding to the non-trivial Berry phase. If this was a classical narrow bandgap semiconductor, the Berry phase should have become a trivial state with the offset value reaching 0.5. In addition, the slope of the Landau fan diagram gives the frequency of SdH oscillation, $F$=5.3 T, corresponding to an extremely small Fermi surface of $S_F = 8\times10^{-5}$ Å$^{-2}$.

Important transport parameters can be obtained from the SdH oscillation analysis.[39] The temperature dependence of the oscillation amplitude $\Delta\rho$ is shown in Fig. 3e, which is described by the formula $\Delta\rho(T) = \Delta\rho(0)\lambda(T)/\sinh(\lambda(T))$. The thermal factor $\lambda(T)$ is given by $\lambda(T) = 2\pi^2 k_B T m^*/\hbar eB$, where $k_B$ is the Boltzmann's constant and $m^*$ is the effective mass. By performing the best fit to the equation, a small effective mass of $m^* = 0.03 m_e$ is derived with $m_e$ being the electron mass. Thus $v_F = 5\times10^5$ m/s can be calculated based on $\hbar k_F = m^* v_F$. Note that the $v_F$ values extracted from the magneto-optics, low-temperature transport and angle-resolved photoemission experiments are of the same order of magnitude but with finite variations, probably due to the electron-phonon interaction[40, 41] or extrinsic effect[42]. We can also estimate the quantum life-time $\tau$ by considering the Dingle factor $e^{-D}$, where $D = 2\pi^2 E_F / \tau e B v_F^2$. Note that $\Delta R / R_0$ is proportional to $[\lambda(T)/\sinh(\lambda(T))]e^{-D}$, so the lifetime can be inferred from the slope in the logarithmic plot (Fig. 3f). By using $m^*=0.03m_e$, the quantum life-time of ZrTe$_5$ can be estimated to be $\tau = 1 \times 10^{-12}$ s. A relatively high quantum mobility can be acquired by $\mu = e\tau/m^* = 7 \times 10^4$ cm$^2$/Vs. Our conclusion in a-c plane is consistent with previous report.[25-27]

As we discussed above, one needs to check different crystal orientations to understand the dimensionality of the Dirac fermions. In order to investigate the overall Fermi surface geometry in ZrTe$_5$, we performed angular-dependent magneto-transport measurements. As shown in Fig. 4a, MR with different field directions is distinguished by colors. A noticeable feature is that all curves overlap with each other when plotted against $B\cos\theta$. At $\theta = 90°$, the longitudinal MR is almost fully suppressed. The close relationship between the MR ratio and out-of-plane magnetic field indicates a 2D-like behavior of the detected orbit. The shape of overall Fermi sphere can be mapped by the angular-dependent SdH oscillations. The negative MR is observed which possibly arises from the axial anomaly.[43] The small amplitude of the negative MR could be attributed to the relatively high Fermi level.[43] Fig. 4b and Fig. 4c demonstrate the extracted SdH oscillations while the magnetic field was tilted from b-axis to a-axis and from b-axis to c-axis, respectively. Similar to the MR background which are overlapped in scale of $B\cos\theta$, the peaks and valleys of the subtracted oscillations are aligned when plotted with respect to $B\cos\theta$, in stark contrast with those isotropic three-dimensional systems[44]. Figure 4d shows a Fast-Fourier-Transform (FFT) analysis of the SdH oscillations. The single peak feature in the FFT spectra corresponds to a single-band transport. For an ideal 2D system, the oscillation frequency $F$ increases linearly with $1/\cos\theta$ and becomes infinitely at $\theta = 90°$ due to the infinite cross-section of Fermi surface. In Fig. 4e and Fig. 4f, we plotted the ideal 2D curve following $1/\cos\theta$ rule. Here, the experimental oscillation frequency $F$ increases with $1/\cos\theta$ and shows a saturation behavior with a small deviations when approaching 90°. The saturation of the frequency suggests a large but not infinite Fermi surface on b-c plane. Combining the trend of the MR background and the SdH oscillations with various field directions, we can conclude that the Fermi surface in $ZrTe_5$ is highly anisotropic and shows long-rod-like shape in the reciprocal space thus possessing a quasi-2D behavior in real space. Here, a single band model is adopted based on the linear Hall effect and single-frequency SdH oscillations at low temperatures (For more detailed discussions, refer to the Section XII of supplementary information).

We have revealed that the Berry phase extracted from the SdH oscillations in a-c plane is exactly $\pi$, corresponding to a topological non-trivial state. However, the same experiments on other orthogonal axes suggest the unexpected trivial state with the complete elimination of Berry phase (Fig. S12). The anomalous feature of Berry phase along different crystal orientations drives us to further analyze the SdH oscillations at all three orthogonal axes for a detailed understanding of band dispersion characteristics. Fig.5a is the oscillation amplitude plotted as a function of temperature. The best fit to those data yields the effective mass for each plane ($k_b$-$k_c$, $k_a$-$k_c$, and $k_a$-$k_b$). The Fermi velocity of each plane can be also calculated as discussed above. The a-c plane shows a much smaller $m^*$ and a larger $v_F$ compared with the other two (Fig. 5b). This value in a-c plane ($5\times10^5$ m/s) is identical to the previous report ($5\times10^5$ m/s)[26] but slightly smaller than those studied by ARPRES ($8\times10^5$ m/s)[25]. As mentioned earlier, the parameters derived from the Landau quantization through transport and magneto-optics are an average value over the closed cyclotron orbit. With the low Fermi level in ZrTe$_5$, the boundary of Brillouin Zone should not have a large impact on the shape of Fermi surface. Thus, we can safely take the Fermi surface as an ellipsoid and the Fermi velocity (effective mass) of each plane is described as, for example, $v_{a-c}=\sqrt{v_a v_c}$. Here, $v_a$ and $v_c$ are the corresponding Fermi velocity along $k_a$ and $k_c$, respectively. The transport parameters along different axes are extracted as shown in Fig. 5c. The $m^*$ (red bar) shows a strong anisotropy with a large enhancement along b-axis. It is natural to observe the relatively small effective mass of $0.02m_e$ & $0.05m_e$ in respectively a-axis and c-axis, consistent with the observation of massless Dirac fermion in a-c plane from our magneto-optics experiments. However, the large effective mass of $1.3m_e$ along b-axis is unexpected for the conventional three-dimensional Dirac materials. Note that the crystal is stacked by the layers on a-c plane along b-axis. The large effective mass may be denoted by the quadratic band dispersion[45] due to the weak interlayer coupling as illustrated in Fig. 5d. Thus, we use "quasi-2D" to describe the Dirac dispersion nature in $ZrTe_5$, which is different from $Sr_2RuO_4$[46] or layered organic superconductors[47] whose Fermi surface is in a warped cylindrical shape and stays open along a certain direction in the Brillouin zone. The much lower Fermi velocity (Fig. 5c) and quantum mobility (Fig. S9) along b-axis also support our assumption. We would like to emphasize that the middle panel in Fig. 5d is a schematic drawing to show the non-Dirac dispersion. An ultra-high resolution APRES measurement along the b-axis direction is required for the detailed dispersion. On the other hand, there is a relatively large difference both in $m^*$ and $v_F$ between a-axis and c-axis, revealing an anisotropy in the in-plane dispersion of Dirac cone. We can use the extracted Fermi velocity along each axis to estimate the overall average Fermi velocity of $6\times10^4$ m/s, which is an order of magnitude lower than that in a-c plane. This value also shows a quantitative agreement with previous zero-field infrared spectroscopy experiments ($3\times10^4$ m/s),[27] where the infrared spectrum without magnetic field includes information along all three axes, thus providing the overall Fermi velocity. Both the low-temperature transport and infrared spectroscopy measurements[27] suggest an incredibly small Fermi velocity along the b-axis and therefore well justify the quasi-2D nature of $ZrTe_5$.

Importantly, the quasi-2D feature of the massless fermions is further evidenced by transport measurements on a high-mobility sample of $5\times10^4$ cm$^2$/Vs, as elaborated in the Supplementary section XIV. The non-trivial Berry phase is not preserved along all the directions in the small magnetic field region, which clearly rules out a 3D Dirac semimetal state. Another important evidence is the observation of the bulk quantum Hall effect (QHE) plateaus with non-trivial Berry phase in a-c plane, which strongly suggests the presence of the quasi-2D Dirac fermions, as also demonstrated in EuMnBi$_2$.[48]

The ab initio calculations have predicted the interlayer binding energy for ZrTe$_5$ to be 12.5 meV/Å, a comparable value to graphite (9.3 meV/Å), suggesting a weak interlayer coupling.[49] Using the conventional scotch tape method, we have also successfully exfoliated ZrTe$_5$ into 2D thin flakes or 1D nano wires (Fig. S10). We found that, upon the exfoliation, not only the inter-plane van der Waals force is easy to overcome, but also that the in-plane Te zigzag chains are readily broken, thus developing a quasi-one-dimensional structure (Fig. S11). The relatively weak bond along c-axis may be the reason for the increase of $m^*$ and the decrease of $v_F$, similar to the case of the b-axis. We also show that the natural 1D structure can be achieved upon exfoliation, which promises related studies such as density wave in ZrTe$_5$ at low dimension.

Since both the magneto-optics and magneto-transport measurements were performed near liquid helium temperature, our conclusion of massless Dirac fermion is valid in the low temperature range. But the resistivity and Hall measurements near 140 K (Fig. S4) reveal a metal-semiconductor transition (anomaly peak) along with Hall sign reversal (Fig. S4), indicating a multi-carrier transport and a complex band structure evolving with temperature. However, for higher temperatures, when thermal energy is larger than the energy level separation ($k_B T > \hbar \omega_c$), Landau quantization is no longer accessible, which calls for further experimental efforts to elucidate the detailed band structure. Both the previous[10] and recent[50] photoemission experiments report the observation of a gap opening near the anomaly temperature. It indicates that the observed Dirac point is unstable against the thermal perturbation. As we discussed in the supplementary section X, the Dirac states that we observed are not likely to originate from the surface states. Recently, several surface sensitive probes were employed to examine the surface/bulk states in ZrTe$_5$.[25, 51] While it was claimed to be gapped with 2D-like structure in the bulk through scanning tunneling microscopy (STM) technique, another study using the same technique realized a gapped topological insulator on the surface layer.[51, 52] Interestingly, an ARPES measurement suggested that the gap opening decreased with reducing the temperature whereas another separate work concluded for massless Dirac fermions at low temperature.[50] The electronic structure was also discussed theoretically. The gapless bulk state was predicted to possibly coexist with the gapped surface state.[33] Importantly, the calculated band structure is highly sensitive to the lattice parameters, which can be tuned by using different growth methods or under specific growth conditions.[53] ZrTe$_5$ is likely to be close to the band topology transition and thus the topology of the band structure is highly sensitive to the chemical composition and the lattice constants. Those theories could explain the controversial arguments surrounding the possible gap. Detailed study on the surface state and clear ARPES experiments along $k_b$ direction are required to comprehensively understand the overall electronic structure.

In the following, we theoretically outline a situation where one can obtain a linear dispersion along two directions and a quadratic one along the third, based on the symmetry analysis of Yang and Nagaosa.[54] For a system with both time reversal and inversion symmetry (which is the case for ZrTe$_5$[53]), an accidental band crossing can be described by the Hamiltonian, $H(k_z, k_x=0, k_y=0) = d_0 + d(k_z, m)\Gamma$. Here $m$ is a tunable parameter, $\Gamma = \tau_z$ or $\Gamma = \sigma_z \tau_z$, $\tau_z$ is the $z$ component of the Pauli matrices describing the orbital pseudospin, while $\sigma_z$ is the $z$ component of the Pauli matrices corresponding to the real spin. A band crossing can be obtained when $d(k_z, m)$ vanishes. If the inversion operator is given by $\tau_z$, then $d(k_z, m)$ is an even function of the crystal momentum $k_z$. Consequently, to the lowest order in $k_z$, $d(\mathrm{k}_z, \mathrm{m}) \approx \mathrm{m} + \frac{1}{2} t_z k_z^2$, where $t_z$ is a constant. By choosing $t_z < 0$, a transition from an insulator to a Dirac semimetal is obtained as $m$ changes sign from negative to positive. Crucially, at the critical point where $m=0$, the band dispersion is quadratic along $k_z$, while it is linear along $k_x$ and $k_y$ (Fig. 5d). Therefore, an energy dispersion in a three-dimensional system, which is quadratic along one direction and linear along the other two, is possible in ZrTe$_5$. The system is at a critical point separating an insulating (or topological insulator) and a 3D Dirac semimetal phase.

The phase transition is possibly driven by changes in the lattice constants.

Judging from the Landau fan diagram in Fig. 3d, the first Landau level can be accessed within 6 T in the magneto-transport which is in stark contrast to the representative Dirac semimetal $Cd_3As_2$ in which the quantum limit requires a magnetic field at least 43 T or even larger[55, 56]. In general, the higher the Fermi level is, the larger field is demanded to access the quantum limit. The calculated Fermi level is 37 meV (refer to Supplementary section VII.) which is an order of magnitude lower than that in $Cd_3As_2$ (typically 200-300 meV). Considering the low Fermi level, the thin flakes of $ZrTe_5$ may be also suitable for static electrical gating, which may add another new degree of freedom to help understanding the anomalous transport behavior along different axes or at high temperatures. While the Dirac states in graphene can be destroyed in the double layers[14], it's of great interest to study the Dirac fermions preserved in $ZrTe_5$ bulk state. Meanwhile, it will be of substantial interests to see whether it will sustain the Dirac fermions as in bulk form or transfer into a quantum spin Hall insulator as theory predicted when approaching the 2D limit. Further experiments in the nanoscale samples or the ultra-quantum limit of $ZrTe_5$ are currently being pursued.

In conclusion, we confirm the existence of Dirac fermions in bulk $ZrTe_5$ by magneto-optics and SdH oscillations. Angular-dependent quantum oscillations reveal a quasi-2D nature of the Dirac fermions, which is further supported by the bulk QHE and the angle-dependent Berry phase at low fields. Theoretical analysis implies that the system is at the critical point between the Dirac semimetal phase and the topological insulator phase. The unusual interlayer dispersion characteristics, low Fermi level, and van der Waals structure demonstrate $ZrTe_5$ as an intriguing system for both the fundamental studies and device applications.

## Method

**Single Crystal synthesis.**

High-quality $ZrTe_5$ single crystals were synthesized by iodine vapor transport method in a two-zone tube furnace. Stoichiometric amounts of high-purity Zr and Te elements were placed in a quartz tube and sealed under vacuum. $ZrTe_5$ is crystallized during a chemical transport reaction process (14 days) with a temperature gradient from 500℃ to 450℃. We use a transport agent of iodine with a concentration of 10 mg cm$^{-3}$. The growth rate along a-axis is much faster than those in b-axis and c-axis, yielding needle-like samples. The chemical ratio is found to be 1:5 for Zr/Te, determined by energy dispersive x-ray analysis in scanning electron microscopy.

**Magneto-optics measurements.**

$ZrTe_5$ was freshly cleaved, subject to an applied magnetic field parallel to the stacking axis (Faraday geometry) at liquid helium temperature. An infrared light from a broadband light source is modulated by interferometer and vibration mirrors in the Fourier-transform infrared spectrometer (FTIR). The light was guided through a light pipe and focused on the sample with millimeter spot by a parabolic cone (Fig. S1). A bolometer was used to measure the intensity of reflection light simultaneously with FTIR.

**Magneto-transport measurements.**

Freshly cleaved $ZrTe_5$ was measured using standard six-terminal Hall-bar geometry in a physical property measurement system. Stanford Research 830 Lock-in amplifiers were used to measure the electrical signals with magnetic field up to 9T that applied for various orientations of the applied magnetic field.

**Sample exfoliation.**

$ZrTe_5$ was firstly exfoliated on scotch tape and then transferred onto 300nm/300μm $SiO_2$/Si substrate. The flakes with different thickness can be identified by different colors. In order to enhance the productivity of one dimensional structure and avoid contamination, PDMS (polydimethylsiloxane) was used to transfer the exfoliated samples.

**Acknowledgements**

This work was supported by the National Young 1000 Talent Plan, the Program for Professor of Special Appointment (Eastern Scholar) at Shanghai Institutions of Higher Learning, National Natural Science Foundation of China (61322407, 11474058), and the Chinese National Science Fund for Talent Training in Basic Science (J1103204). Part of the sample fabrication was performed at Fudan Nano-fabrication Laboratory. A portion of this work was performed at the National High Magnetic Field Laboratory, which is supported by National Science Foundation Cooperative Agreement No. DMR-1157490 and the State of Florida.

**Author contributions**

F.X. conceived the idea and supervised the experiments. X.Y. and C.Z. carried out the growth, magneto-transport measurements and data analysis. X.Y. and C.Z. performed the magneto-optics experiments. Y.L., S.S., X.S. and J.Z. helped with the growth. Y.L., C.S., and H.Y. helped with the data analysis. A.N. and S.S. performed the band structure modeling. J.X., H.Y. and Z.A. assisted in the infrared property characterizations. X.Y., C.Z. and F.X. wrote the paper with helps from all other authors.

**FIGURES**

**Figure 1 | Landau-level fan chart and normalized magneto-reflection of ZrTe$_5$. a**, Landau levels in ZrTe$_5$ as a function of magnetic field. Arrows show the allowable optical transitions. Different colors denote different resonance energy. **b**, The resonance energy versus $\sqrt{n}+\sqrt{n+1}$ under different magnetic fields. The straight line is for eye-guide. Inset: Normalized reflection spectra under different magnetic fields. Triangles denote the peak positions.

**Figure 2 | Magneto-reflectivity in ZrTe$_5$.** Reflectivity maxima frequency plotted with $\sqrt{B}$. The observed inter-Landau-level resonance clearly follows a $\sqrt{B}$ dependence. Inset is the reflectivity maxima frequency versus $B$.

**Figure 3 | Magneto-transport measurements in ZrTe$_5$. a**, Schematic drawing of the transport measurements. A constant current was applied along a-axis. **b**, Angular dependence of MR at 2.5 K. The SdH oscillations are observed at different angles. **c**, MR at different temperatures with $\theta=0°$. The oscillations are not observable above 20 K. The inset is an enlarged view of the oscillations. **d**, Landau fan diagram at $\theta=0°$. Non-trivial Berry phase can be clearly determined. **e**, Temperature dependence of the SdH oscillation amplitude at 1.34 T. **f**, Dingle plot for the extraction of the scattering time.

**Figure 4 | Angular-dependent Magneto-transport measurements in ZrTe$_5$. a**, Resistivity versus magnetic field. Data for different $\theta$, denoted by different colors, overlap with each other, indicating a quasi-2D behavior. **b** and **c**, The oscillatory component $\Delta\rho$ plotted against $1/B\cos(\theta)$ or $1/B\cos(\phi)$ for $B$ rotation along c-axis and a-axis. **d**, FFT spectra for different $\theta$. **e** and **f**, Oscillation frequency plotted with angle. The black curve shows an ideal 2D behavior.

**Figure 5 | Dispersion characteristics along orthogonal axes. a**, The temperature dependence of the SdH oscillation amplitude, for magnetic field parallel to c-axis (upper), a-axis (middle) and b-axis (lower). **b**, The effective mass and Fermi velocity retracted from the SdH oscillations in different crystal planes. **c**, The effective mass and Fermi velocity calculated from b, based on the ellipsoid model. **d**, A schematic drawing for dispersion characteristics along different orientations. Dirac dispersion is found along a and c axes and non-linear dispersion along the b-axis. The detail $k_b$ dispersion needs to be investigated using high-resolution ARPES spectroscopy.

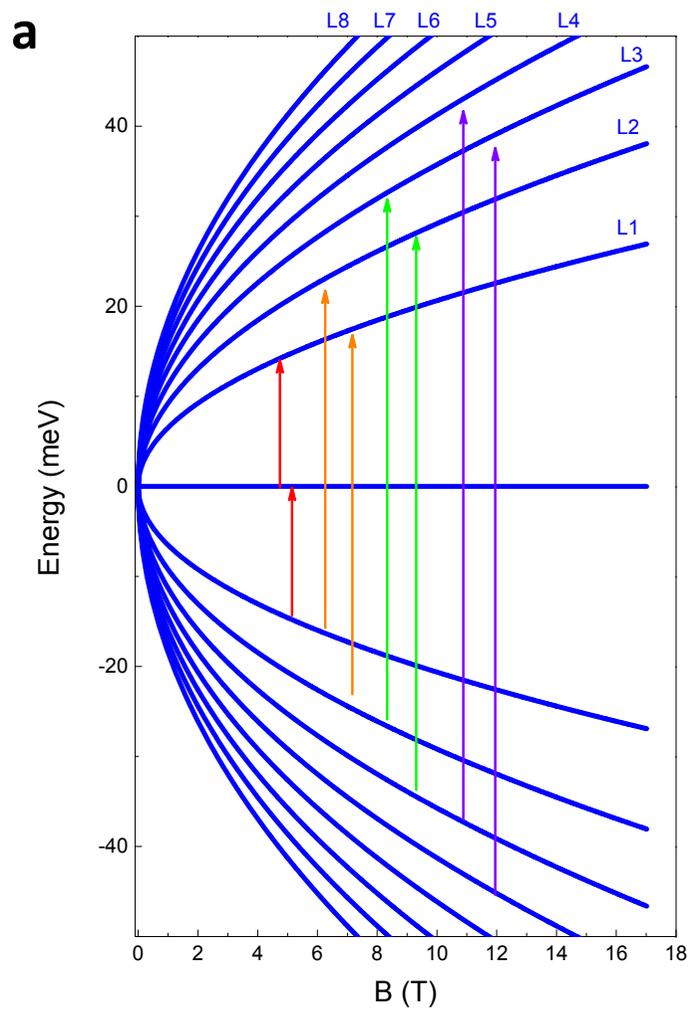 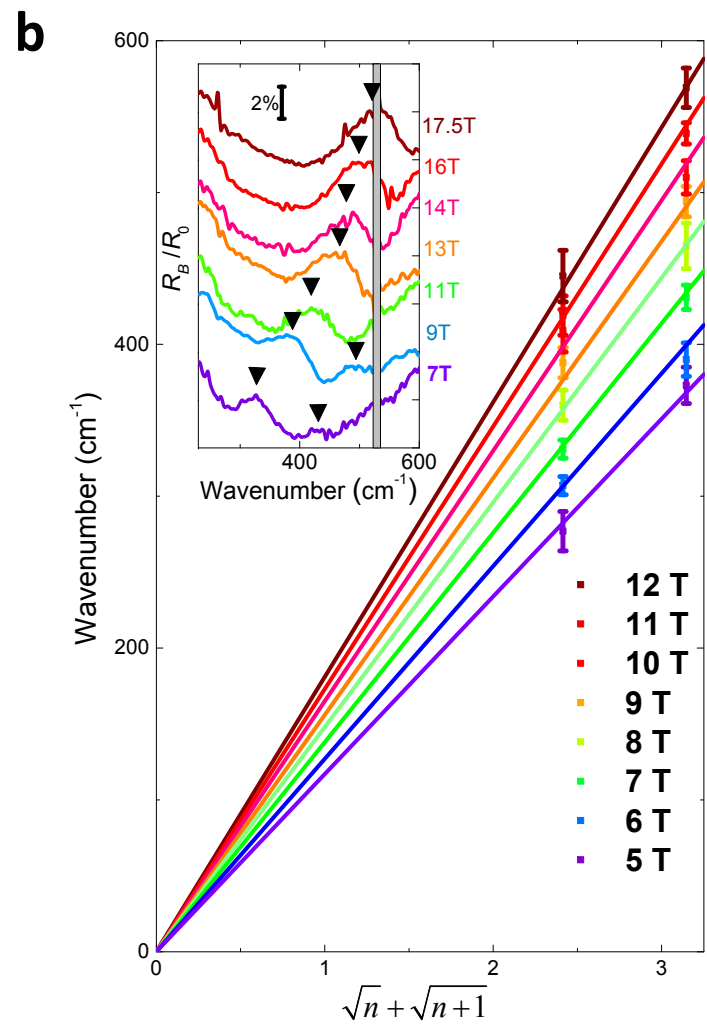

X. Yuan, et al. Figure 1

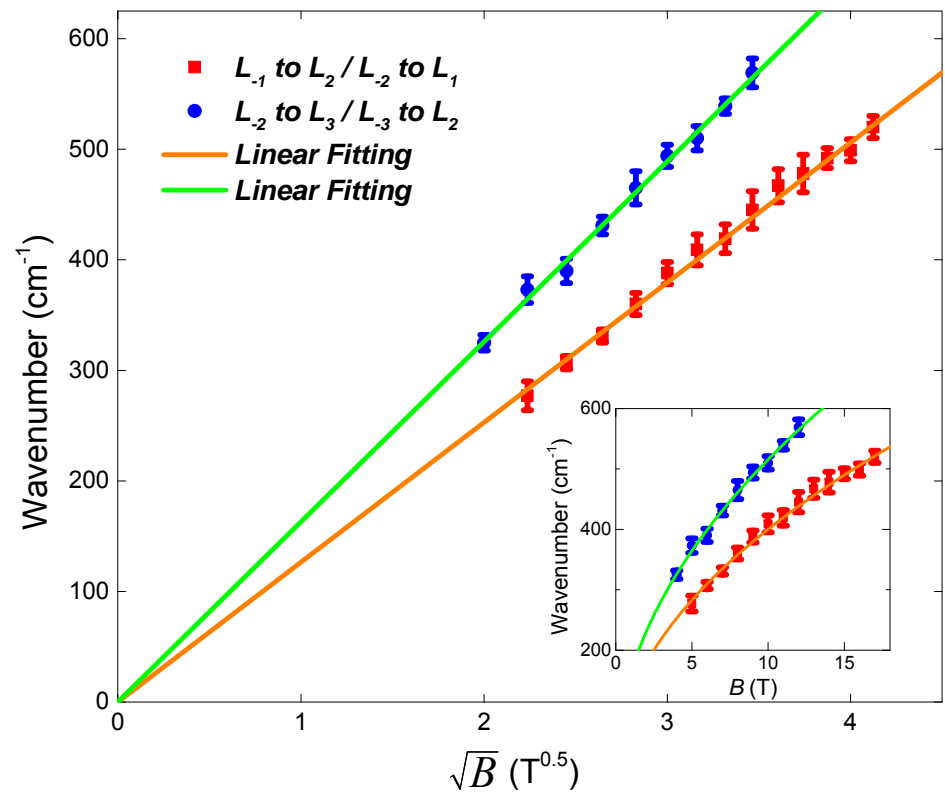

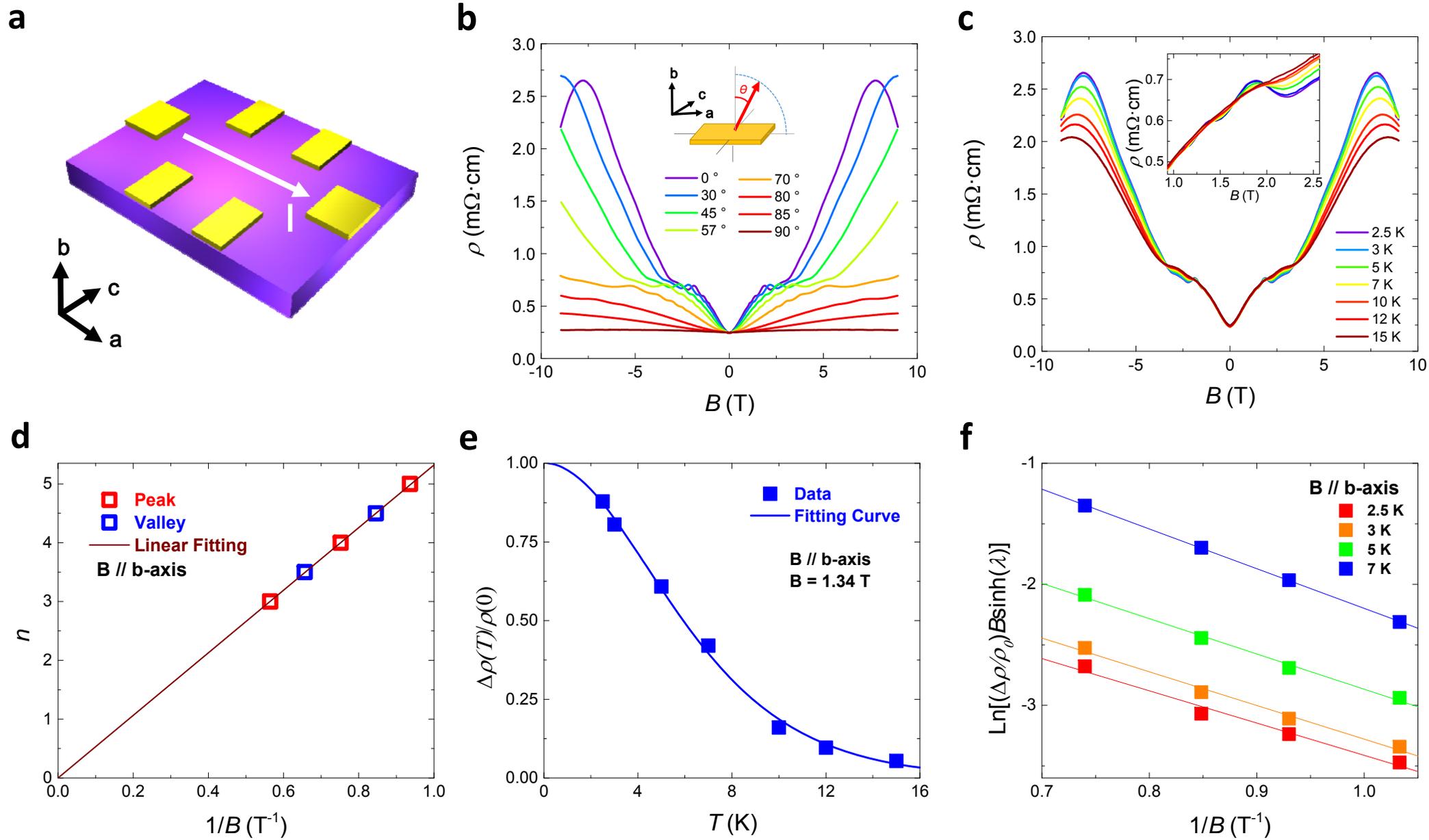

X. Yuan, et al. Figure 3

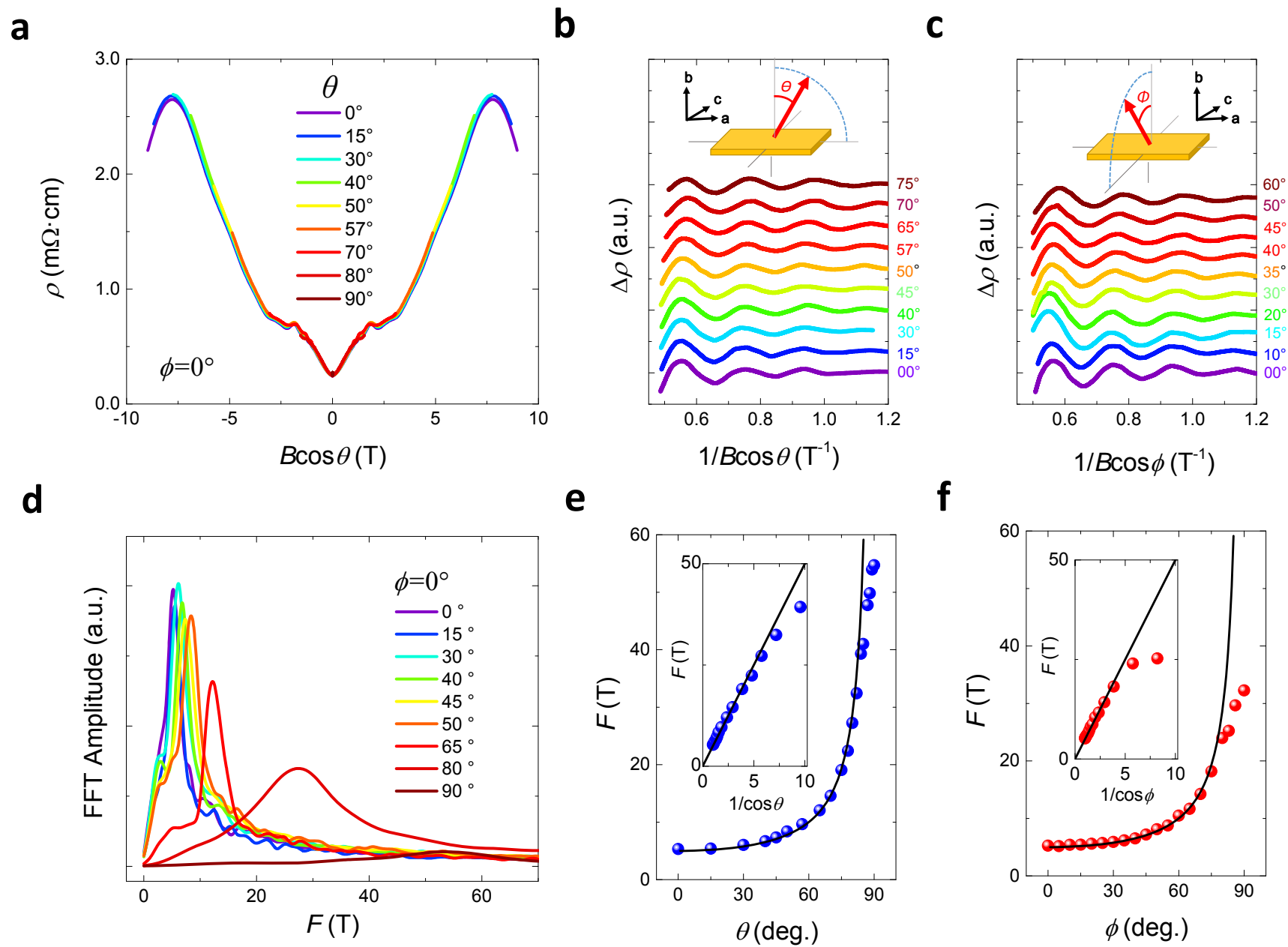

X. Yuan, et al. Figure 4

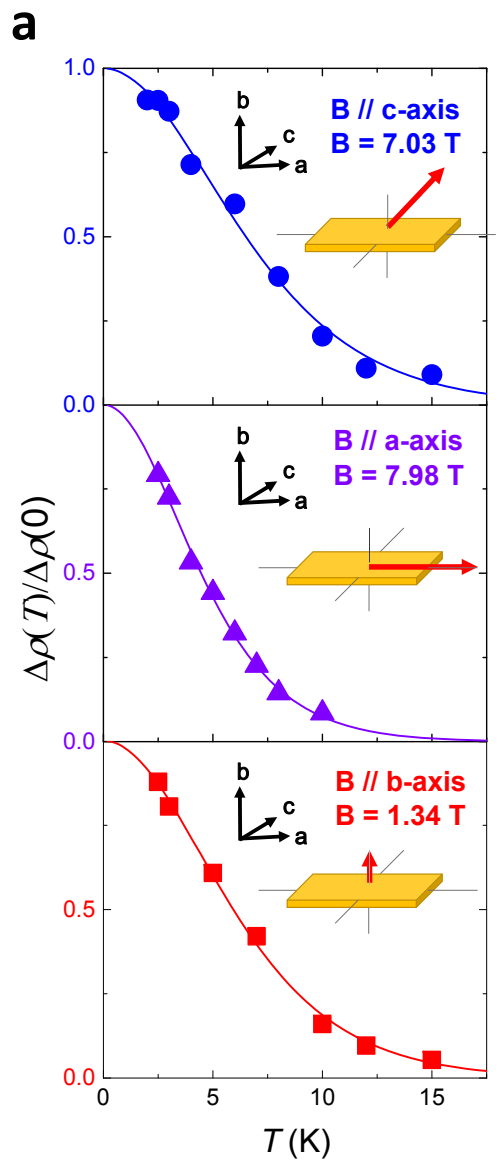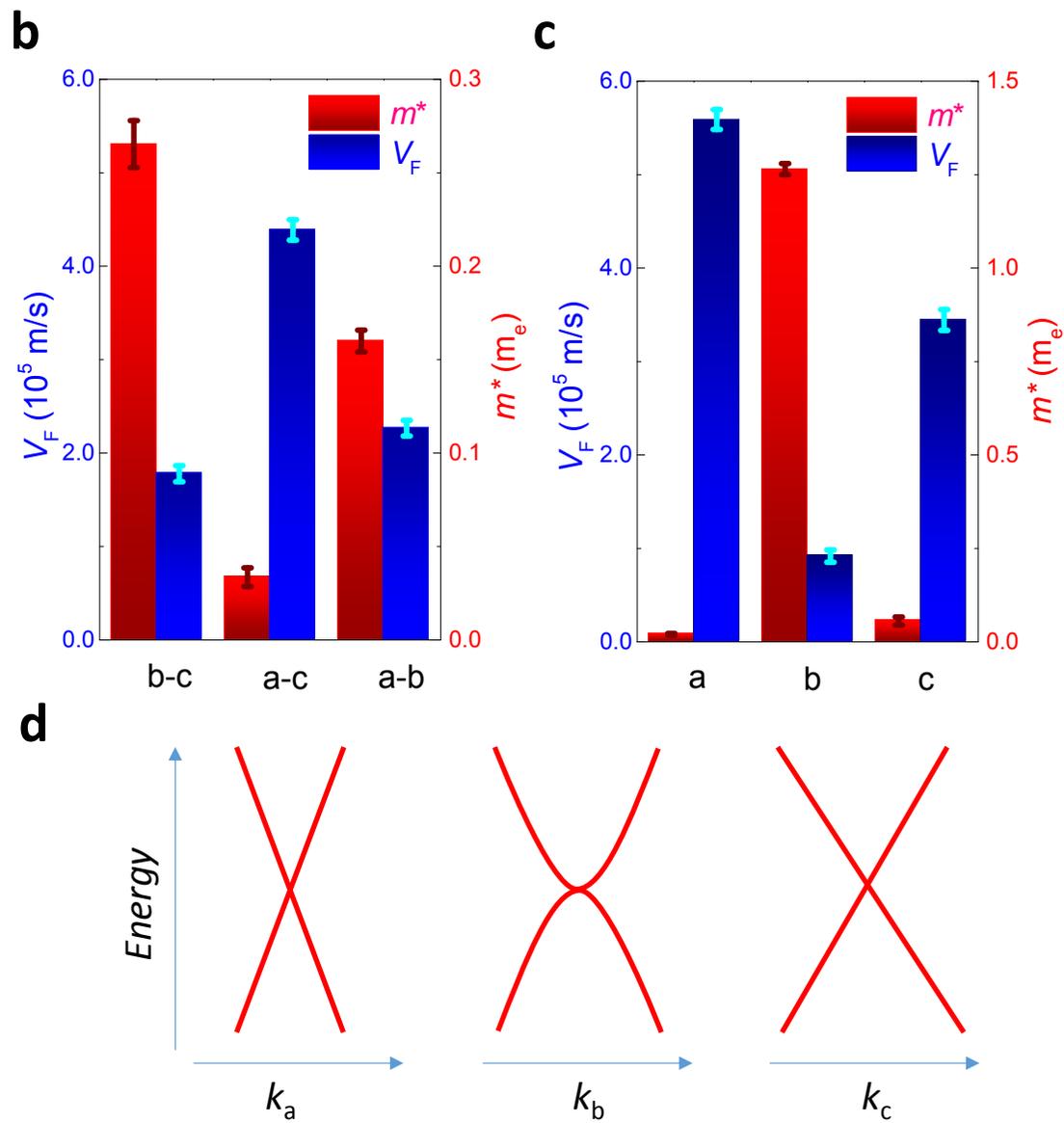

X. Yuan, et al. Figure 5